\documentclass[prl,preprint,showpacs]{revtex4}

\begin{document}
\author{James W. Dufty}
\affiliation{Department of Physics, University of Florida,
Gainesville, Florida 32611}
\author{J. Javier Brey}
\affiliation{F\'{\i}sica Te\'{o}rica, Universidad de Sevilla,
E-41080 Sevilla, Spain}
\title{\textbf{Hydrodynamic Modes for Granular Gases}}
\date{\today }
\begin{abstract}
The eigenfunctions and eigenvalues of the linearized Boltzmann
equation for inelastic hard spheres ($d=3$) or disks ($d=2$)
corresponding to $d+2$ hydrodynamic modes, are calculated in the
long wavelength limit for a granular gas. The transport
coefficients are identified and found to agree with those from the
Chapman-Enskog solution. The dominance of hydrodynamic modes at
long times and long wavelengths is studied via an exactly solvable
kinetic model. A collisional continuum is bounded away from the
hydrodynamic spectrum, assuring a hydrodynamic description at long
times. The bound is closely related to the power law decay of the
velocity distribution in the reference homogeneous cooling state.

\end{abstract}
\maketitle

The simplest model for a granular gas at low density is given by
the Boltzmann equation for smooth, inelastic hard spheres or disks
\cite {vNyE01}. It provides an appropriate context in which to
address a number of fundamental issues. Primary among these is the
existence of a macroscopic fluid dynamics analogous to the
Navier-Stokes description for real gases. The derivation of
hydrodynamic equations from the inelastic Boltzmann equation and
identification of expressions for the transport coefficients has
been a problem of interest for two decades
\cite{JyR85,LSJyCh84,GyS95,SyG98}.

Accurate prediction of the transport coefficients as a function of
the restitution coefficient \cite{BDKyS98}, and confirmation via
Monte Carlo simulation \cite{ByC01}, has been accomplished only
within the last few years. However, the method used in these
derivations (Chapman-Enskog) is formal and has been susceptible to
debate. It is of interest therefore to look for a hydrodynamic
description in the simplest context of linear response to a small
perturbation of the ``universal'' homogeneous cooling state (HCS).
This response is provided by the linearized Boltzmann equation,
and the existence of hydrodynamics depends on the spectrum of the
operator. Another issue is the isolation of the hydrodynamic
spectrum from the remaining ``microscopic'' excitations, such that
the hydrodynamic modes dominate on the long space and time scales.
For elastic collisions, this has been proven for ``hard''
potentials (including hard spheres) \cite{Mc89}.

Here a calculation of the hydrodynamic spectrum from the
linearized inelastic Boltzmann equation is reported. The transport
coefficients are identified and shown to agree with those from the
Chapman-Enskog method. Moreover, some support for the isolation of
the hydrodynamic spectrum is provided by an exact analysis of a
model Boltzmann equation \cite{BMyD96,BDKyS98}. The model exhibits
the dominant features of granular gases, including a non-trivial
HCS with algebraic decay at large velocities, discussed
extensively for related Maxwell models \cite{EyB02}. It is shown
that the pure point spectrum for elastic collisions develops a
continuum for any degree of inelasticity, extending toward the
hydrodynamic spectrum. However, it remains bounded away from the
latter by an amount that is controlled by the power law of the
decay of the velocity distribution. This relationship of the
spectrum to the reference HCS is unexpected, but it is argued that
it is more general than the limitations of the model.

The inelastic nonlinear Boltzmann equation for the density
$f\left( \mathbf{r},\mathbf{ v,}t\right)$ of particles of mass $m$
at position $\mathbf{r}$ with velocity $\mathbf{v}$ at time $t$
has the form
\begin{equation}
\left( \partial _{t}+\mathbf{v\cdot \nabla }\right) f=J[f,f],
\label{1}
\end{equation}
where $J[f,f]$ is a bilinear functional of $f$ \cite{vNyE01}.
There is no stationary solution to this equation for an isolated
system. However, a homogeneous cooling solution (HCS) is assumed
to exist with a scaling property similar to the Maxwellian,
\begin{equation}
f_{HCS}(\mathbf{v},t)=n_{H}v_{0}^{-d}(t)\chi_{0}(\mathbf{c}),\quad
\mathbf{c}=\frac{\mathbf{v}}{v_{0}(t)}\, . \label{2}
\end{equation}
Here $n_{H}$ is the homogeneous density and
$v_{0}(t)=\sqrt{2T(t)/m}$ is the thermal velocity defined in terms
of the temperature $T(t)$. The latter obeys the equation
$T^{-1}\partial_{t} T=-\zeta_{0} $, where the cooling rate
$\zeta_{0}=\zeta[ \chi _{0},\chi _{0}]$ also is a specified
bilinear functional. Both $\chi_{0}$ and $\zeta_{0}$ must be
determined self-consistently from the Boltzmann equation
\begin{equation}
\frac{1}{2}\zeta_{0} \frac{\partial}{\partial {\bf c}} \cdot
\left( {\bf c} f_{HCS} \right)=J[f_{HCS},f _{HCS}]. \label{4}
\end{equation}
The existence of $f_{H}$ is supported by approximate polynomial
expansions \cite{vNyE98} and by Monte Carlo simulation
\cite{BMyC96}.

To study the relaxation of small spatial perturbations of the HCS,
$\Delta$ is defined by $f=f_{HCS} \left[ 1+\Delta \right]$, and
terms up through linear order in $\Delta $ are retained in the
Boltzmann equation. The resulting equation is then written in
dimensionless form with the velocities scaled relative to
$v_{0}(t)$ and the time expressed in terms of
$s=\int^{t}dt^{\prime }\nu_{0} \left( t^{\prime }\right) $, where
$\nu_{0}(t) $ is the average collision frequency, proportional to
$\sqrt{T(t)}$. It is sufficient to consider a single Fourier
component $\tilde{\Delta}(\mathbf{k})$ that obeys
\begin{equation}
\left( \partial _{s}-i\mathbf{k\cdot c}+\mathcal{L}\right)
\tilde{\Delta} =0, \label{5}
\end{equation}
\begin{eqnarray}
\mathcal{L}\tilde{\Delta}&=&-\chi _{0}^{-1}\left\{ J[\chi
_{0},\chi _{0}\tilde{\Delta} ]+J\left[ \chi _{0}\tilde{\Delta}
,\chi _{0}\right] \right. \nonumber \\
&& \left. -\frac{\zeta_{0}}{2} \frac{\partial }{\partial
\mathbf{c}}\cdot \left( \mathbf{c}\chi _{0}\tilde{\Delta} \right)
\right\} . \label{6}
\end{eqnarray}
The solution is sought in a Hilbert space defined by the scalar
product
\begin{equation}
\left( a,b\right) =\int d\mathbf{c}\, \chi _{0}({\bf c})
a^{\ast}({\bf c}) b({\bf c}). \label{7}
\end{equation}
The relevant eigenvalue problem is
\begin{equation}
\left( -i\mathbf{k\cdot c}+\mathcal{L}\right) \phi _{i}=\omega
_{i}\left( k\right) \phi _{i}.  \label{8}
\end{equation}
It is understood that the index $i$ may be discrete or continuous.
The hydrodynamic spectrum can be defined precisely as follows. The
linearized Boltzmann equation provides exact balance equations for
the moments of $\Delta $ corresponding to the density, flow
velocity, and temperature. The latter are the $d+2$ hydrodynamic
fields. The spectrum of these balance equations can be calculated
without approximation in the limit $\mathbf{k}=0$. For elastic
collisions, there is a $(d+2)$-fold degenerate point at zero
eigenvalue, corresponding to the conservation laws for binary
collisions. Their perturbation at finite $\mathbf{k}$ defines the
hydrodynamic modes more generally. For inelastic collisions, the
spectrum is again $d+2$ points, now at $0,\zeta_{0} /2 ,-\zeta_{0}
/2$, with the latter being $d$-fold degenerate. Again, the
hydrodynamic modes more generally are defined as those solutions
to (\ref{8}) that are continuously connected as functions of $k$
to these special $k=0$ solutions. For elastic collisions, it is
possible to prove analyticity about $k=0$ \cite{Mc89}, and it is
assumed this is the case here as well.

The first step is to show that $\mathcal{L}$ includes the spectrum
of the balance equations at $k=0$, and to determine the
corresponding eigenfunctions. For elastic collisions, the
stationary state is given by $J[f_{M},f_{M}]=0$, whose solution is
the Maxwellian. Differentiating this stationary equation with
respect to the hydrodynamic fields leads to the linearized
Boltzmann operator acting on combinations of $1,\mathbf{c},c^{2}$
independently being zero. These are the $d+2$ eigenfunctions with
zero eigenvalue. A similar analysis works for the inelastic case,
using the stationary condition (\ref{4}). Straightforward
calculations give
\begin{equation}
\mathcal{L}\phi _{i}^{(0)}=\omega _{i}^{(0)}\phi _{i}^{(0)},\quad
i=1,..,d+2,  \label{9}
\end{equation}
\begin{equation}
\phi^{(0)} \rightarrow \left\{ d+1+{\bf c}\cdot \frac{\partial \ln
\chi_{0}}{\partial {\bf c}}, -d-{\bf c}\cdot \frac{\partial \ln
\chi_{0}}{\partial {\bf c}},
 -\frac{\partial \ln
\chi_{0}}{\partial {\bf c}} \right\},
 \label{10}
\end{equation}
\begin{equation}
\omega _{i}^{(0)} \rightarrow \left\{ 0,\zeta/2,-\zeta/2 \right\}.
\label{11}
\end{equation}
The evaluation of the hydrodynamic spectrum at finite $k$ is now a
well-defined technical problem. In general, the eigenvalues depend
on both $k$ and the restitution coefficient $\alpha $. It is known
from the Navier-Stokes equations that the behavior for small $k$
$\left( \alpha \right) $ is not uniform with respect to $\alpha $
$\left( k\right) $, so that a perturbation expansion in one or the
other does not provide the entire mode structure. Here, for
illustration, only the results for small $k$ are reported. The
expansion is defined by $\omega_{i}\left( k\right)
=\omega_{i}^{(0)}+ik\omega_{i}^{(1)}+k^{2}\omega
_{i}^{(2)}+\cdots$, and $\phi _{i}\left( k\right)
=\phi_{i}^{(0)}+ik\phi_{i}^{(1)}+k^{2}\phi _{i}^{(2)}+\cdots$. The
leading terms are those of Eqs.\ (\ref{10}) and (\ref{11}). Since
$\mathcal{L}$ is not self-adjoint, it is necessary to introduce a
set of functions $\left\{ \psi _{i}\right\} $ that are
biorthogonal to $\left\{ \phi _{i}\right\} $. The leading terms of
a similar expansion in $k$ are found to be
\begin{equation}
\psi_{i}^{(0)} \rightarrow \left\{ 1,
\frac{c^{2}}{d}+\frac{1}{2},\widehat{\mathbf{k}}\cdot \mathbf{c},
\widehat{\mathbf{e}} ^{(i)}\cdot \mathbf{c} \right\}, \label{11a}
\end{equation}
where $\{ \widehat{\mathbf{k}}, \widehat{\mathbf{e}}^{(i)} \}$ are
$d$ pairwise orthogonal unit vectors. The results to second order
in $k$ are $\omega_{i}^{(1)}=0$,
\begin{equation}
\omega _{i}^{(2)}=\left(
\psi_{i}^{(0)},\widehat{\mathbf{k}}\mathbf{\cdot c}\phi
_{i}^{(1)}\right) +\delta _{i2} \zeta_{0} \left[ \chi
_{0},\chi_{0} \mathcal{Q}_{1} \phi _{2}^{(2)} \right],  \label{14}
\end{equation}
\begin{equation}
\phi _{i}^{(1)}=\mathcal{Q}_{i}\left( \mathcal{L}-\omega
_{i}^{(0)}\right) ^{-1}\widehat{\mathbf{k}}\mathbf{\cdot c}\phi
_{i}^{(0)}, \label{15}
\end{equation}
\begin{equation}
\phi _{i}^{(2)}=-\mathcal{Q}_{i}\left( \mathcal{L}-\omega
_{i}^{(0)}\right)^{-1}\widehat{\mathbf{k}}\mathbf{\cdot c}\phi
_{i}^{(1)}. \label{16}
\end{equation}
The operators $\mathcal{Q}_{i}=1-\mathcal{P}_{i}$ are projections
orthogonal to $\phi _{i}^{(0)}$ with $\mathcal{P}_{i}a=\phi
_{i}^{(0)}\left( \psi _{i}^{(0)},a \right)$. The constants $\omega
_{i}^{(2)}$ are transport coefficients which are identified from a
similar calculation using the Navier-Stokes equations. For
example, the shear viscosity $\eta $ is
\begin{equation}
\eta =\frac{m v_{0}\omega_{4}^{(2)}}{\sigma^{d-1}}=-\frac{m
v_{0}}{\sigma ^{d-1}} \left( c_{x}c_{y},\left(
\mathcal{L}+\frac{1}{2}\zeta_{0} \right) ^{-1}c_{x} \frac{\partial
\ln \chi_{0}}{\partial c_{y}}\right) . \label{17}
\end{equation}
This agrees with the result obtained from the Chapman-Enskog
method \cite{DyB02}. A more complete connection with the latter
can be established as follows. Assuming that the hydrodynamic
spectrum dominates for long times and small $k$, the solution to
the Boltzmann equation for small perturbations becomes
\begin{equation}
f\rightarrow f_{hcs}\left( \mathbf{v}\right) \left[
1+\sum_{i=1}^{d+2}\phi _{i}\left( k\right) \delta y_{i}^{h}\left(
\mathbf{k},s\right) \right] . \label{18}
\end{equation}
Here $\delta y_{i}^{h}\left( \mathbf{k},s\right) $ are the
hydrodynamic fields
\begin{equation}
\delta y_{i}^{h}\left( \mathbf{k},s\right) =\left( \psi
_{i},\tilde{\Delta} \left( \mathbf{k},s\right) \right) =e^{s\omega
_{i}\left( k\right) }\left( \psi _{i},\tilde{\Delta} \left(
\mathbf{k},0\right) \right).   \label{19}
\end{equation}
The connection between these fields and the macroscopic density,
temperature, and flow velocity is obtained for small $k$ by using
$\psi _{i}\rightarrow \psi _{i}^{(0)}$ in (\ref{19}). The fields
$\delta y_{i}^{h}\left( \mathbf{k} ,s\right) $ represent
components of the initial distribution that obey the hydrodynamic
equations for all times, whereas the density, temperature, and
flow velocity become equal to these only on the long space and
time scales \cite{Mc89}. Use of $\phi _{i}\left( k\right)
\rightarrow \phi _{i}^{(0)}+ik\phi _{i}^{(1)}$ for small $k$ \ in
(\ref{18}) leads to a result that is in agreement with the
Navier-Stokes approximation of the Chapman-Enskog method
\cite{BDyR03}.

The existence of a  hydrodynamic spectrum is relevant only if the
corresponding modes dominate for long times and long wavelengths.
A sufficient condition is that the magnitude of the hydrodynamic
eigenvalues be smaller than that of all other parts of the
spectrum. This is the case for elastic collisions, but the proof
does not extend directly to $-i\mathbf{ k\cdot c}+\mathcal{L}$
here and the question remains open in general \cite{G01}. In fact,
the mode at $\zeta_{0} /2$ gives reason for concern since it might
become comparable to the microscopic excitations at large
inelasticity. To explore this point, it is instructive to consider
a model for the Boltzmann equation that preserves the essential
physics of a non-trivial HCS, exact balance equations, and
hydrodynamic modes, as described above. The Boltzmann collision
operator can be written
\begin{equation}
J[f,f]=-\nu \left( f-g\right),   \label{20}
\end{equation}
where $\nu f$ is the ``loss'' contribution and $\nu g$ is the
``gain'' contribution. Both the collision frequency $\nu $ and $g$
are specified positive functionals of $f$.

The class of models to be considered here are characterized by the
following two simplifications: 1) $\nu $ is independent of
$\mathbf{v}$ and, 2) $\nu $ and $g$ are functionals of $f$ only
through its moments with respect to $1,\mathbf{v},v^{2}$ . The
first condition is similar to that of ``Maxwell models'' that have
been introduced recently \cite{EyB02}. The second condition means
that $\nu $ and $g$ depend on the state of the system only through
the hydrodynamic fields. The forms of these functionals are
subjected to preserve the exact properties of the collision
operator required by the balance equations,
\begin{equation}
\int d\mathbf{v}\left(
\begin{array}{c}
1 \\
\mathbf{v} \\
mv^{2}
\end{array}
\right) g\left( \mathbf{r},\mathbf{v},t\right) =\left(
\begin{array}{c}
n \\
n\mathbf{u} \\
nTd \left( \lambda ^{2}+\frac{mu^{2}}{Td} \right)
\end{array}
\right).   \label{21}
\end{equation}
Here ${\bf u}$ is the flow velocity and $\lambda ^{2}\equiv
1-\zeta /\nu $. In principle, $\zeta$ is the same functional of
$f$ as obtained from the Boltzmann equation, although it can be
eventually approximated similarly to $\nu$ and $g$. I any case,
Eq.\ (\ref 21) must be preserved. In the following, we will assume
that both $\zeta$ and $\nu$ scale as $T^{1/2}$  in order to mimic
hard sphere behavior. With the above class of models, it is
obtained from Eq.\ (\ref {4})
\begin{equation}
\chi _{0}\left( \mathbf{c}\right) =\int_{1}^{\infty }dxP(x)\left(
\lambda x\right) ^{-d}g^{\ast }\left( \mathbf{c}/\lambda x\right),
\label{23}
\end{equation}
where  $P(x)\equiv px^{-\left( 1+p\right) }$ with $p=2\nu /\zeta
$. The dimensionless gain function $g\left( \mathbf{v};y^{h}({\bf
r},t)\right)=n_{H}\left( \lambda v_{0}\right) ^{3}g^{\ast }\left(
\mathbf{c}/\lambda \right) $ has been introduced. Further
investigation requires specification of $ g^{\ast }\left(
\mathbf{c}/\lambda \right) $. However, it is easily verified that
moments of degree equal or greater than $p$ are not finite if the
corresponding moments of $g^{\ast }$ exist. This is consistent
with a similar result for the Maxwell models where algebraic decay
of $\chi _{0}\left( \mathbf{c} \right) $ is observed for large
velocities, also due to a slower decay of the loss term than the
gain term.

Linearization of the kinetic model around this HCS leads again to
Eq. (\ref {5}), but with a simpler form for $\mathcal{L}$. It has
a decomposition into the subspace of $\phi _{i}^{(0)}$ and its
complement, $\mathcal{L}=\mathcal{PLP}+\mathcal{QLQ}$, where
$\mathcal{P}=\sum \mathcal{ P}_{i}$ and
$\mathcal{Q}=1-\mathcal{P}$. In the subspace of $\phi_{i}^{(0)}$
the spectrum is just the $d+2$ points of (\ref{11}),
\begin{equation}
\mathcal{PLP}=\sum_{i=1}^{d+2}\omega _{i}^{(0)}\mathcal{P}_{i}.
\label{25}
\end{equation}
In the complementary subspace the operator is
\begin{equation}
\mathcal{QLQ}=\nu \mathcal{Q}+\chi _{0}^{-1}\frac{\zeta_{0}}{2}
\frac{
\partial }{\partial \mathbf{c}}\cdot \left( \mathbf{c}\chi _{0}\mathcal{Q}
\right).   \label{26}
\end{equation}
Interestingly, there is no explicit dependence on the gain
function $g^{\ast }$ except through $\chi _{0}$. For elastic
collisions, the second term is missing and the model becomes the
linearized Bhatnager-Gross-Krook model \cite{BGK75}, with a single
degenerate point in the spectrum representing all the microscopic
excitations of the Boltzmann operator. It is somewhat easier to
analyze the corresponding adjoint problem
\begin{equation}
\nu \phi -\frac{1}{2}\zeta_{0} \mathbf{c}\cdot \frac{\partial
}{\partial \mathbf{ c}}\phi =\omega \phi ,\quad \phi
=\mathcal{Q}^{\dagger }\phi, \label{27}
\end{equation}
where $\mathcal{Q}^{\dagger}$ is the adjoint of $\mathcal{Q}$.
Solutions to this equation exist from the class of homogeneous
functions
\begin{equation}
\phi \propto \prod_{i=1}^{d} c_{i}^{q_{i}}, \quad \sum_{i=1}^{d}
q_{1}=q. \label{28}
\end{equation}
The property $\mathcal{Q}^{\dagger }\phi =\phi $ can be satisfied
with a suitable choice of the linear combinations. The
corresponding eigenvalue is $ \omega =\nu -q\zeta_{0} /2$. Thus
the spectrum of $\mathcal{L}$ includes a continuum with $\Re
\omega \leq \nu $. Of critical interest for hydrodynamics  is
whether the lower bound for this continuum on the real axis
intersects the hydrodynamic point spectrum. The bound is
determined by the maximum degree of such homogeneous
eigenfunctions admitted in the Hilbert space. The condition is
that $\left\| c^{q}\right\| ^{2}=\left( c^{q},c^{q}\right)$
exists. Using Eq.\ (\ref{23}), this becomes
\begin{equation}
\left\| c^{q}\right\| ^{2} =\lambda ^{2\Re q}p
M_{q}\int_{1}^{\infty }dxx^{2\Re q-\left( 1+p\right) },
\label{29}
\end{equation}
\begin{equation}
 M_{q} =\int_{0}^{\infty }d{\bf c} c^{2 \Re q}g^{\ast
}\left( c\right). \label{30}
\end{equation}
For finite $M_{q}$, the value of $q$ is bounded from above by $\Re
q<q_{m}=p/2=\nu /\zeta_{0}$. The continuous spectrum therefore is
restricted by $\nu /2< \Re \omega \leq \nu $. This means that the
discrete spectrum is isolated from the continuum for all
$\zeta_{0} /\nu <1$.  But $\zeta_{0} /\nu $ must be smaller than
unity because of the third condition required by Eq.\ ({\ref{21}).
This condition is related with the fact that only a fraction of
the energy of the particles can be lost in collisions. The
continuum is singular in the sense that it is independent of the
restitution coefficient but it is present only for finite
inelasticity.

For this class of models, it is seen that the hydrodynamic
spectrum at $k=0$ is isolated from the rest of the spectrum. The
model also allows exact calculation of the hydrodynamic
eigenvalues $\omega _{i}\left( k\right) $ at finite $k$, without
the restrictions of the perturbation theory calculation above. For
elastic collisions, such a calculation shows that the hydrodynamic
modes exist only for $k$ less than some maximum value. It is
expected that a similar condition applies in the granular case as
well.

 The research of J.W.D. was supported by Department of Energy
Grants DE-FG03-98DP00218 and DE-FG02ER54677. The research of
J.J.B. was partially supported by the Ministerio de Ciencia y
Tecnolog\'{\i}a (Spain) through Grant No. BFM2002-00307.

\end{document}